\documentstyle[12pt,epsf]{article}

\begin{document} 

\begin{titlepage}

\hrule 
\leftline{}
\leftline{Chiba Univ. Preprint
          \hfill   \hbox{\bf CHIBA-EP-120}}
\leftline{\hfill   \hbox{hep-th/0004158}}
\leftline{\hfill   \hbox{April 2000}}
\vskip 5pt
\hrule 
\vskip 1.0cm
\centerline{\large\bf 
  Abelian dominance in low-energy Gluodynamics
} 
\vskip 0.5cm
\centerline{\large\bf  
due to dynamical mass generation$^*$ 
}
\centerline{\large\bf  
}

\vskip 1cm

\centerline{{\bf 
Kei-Ichi Kondo$^{1,2,}{}^{\dagger,}{}^*$
and Toru Shinohara $^2{}^{\ddagger}$
}}  
\vskip 1cm
\begin{description}
\item[]{\it \centerline{ 
$^1$ Department of Physics, Faculty of Science, 
Chiba University,  Chiba 263-8522, Japan}
  }
\item[]{\it 
$^2$ Graduate School of Science and Technology,
  Chiba University, Chiba 263-8522, Japan
  }
\end{description}

\centerline{{\bf Abstract}} 
We show that off-diagonal gluons and off-diagonal ghosts acquire their masses dynamically in QCD if the maximal Abelian gauge is adopted.  This result strongly supports the Abelian dominance in low-energy region of QCD.    The mass generation is shown to occur due to ghost--anti-ghost condensation caused by attractive quartic ghost interactions within the Abelian projected effective gauge theory (derived by one of the authors).  In fact, the quartic ghost interaction is indispensable for the renormalizability due to nonlinearity of the maximal Abelian gauge.   
 The ghost--anti-ghost condensation is associated with the spontaneous breaking of global $SL(2,R)$ symmetry recently found by Schaden at least for SU(2) case.  
Moreover we write down a new extended BRS algebra in the maximal Abelian gauge 
which should be compared with that of Nakanishi-Ojima for the Lorentz gauge.  Finally, we argue that the mass generation may be related to the spontaneous breaking of a supersymmetry $OSp(4|2)$ hidden in the maximal Abelian gauge.


\vskip 0.2cm
\hrule  
$^\dagger$ 
  E-mail:  kondo@cuphd.nd.chiba-u.ac.jp 
\par 
$^\ddagger$ 
  E-mail: sinohara@cuphd.nd.chiba-u.ac.jp
\par\noindent


\vskip 0.5cm  

$^*$ To be published in Phys. Lett. B.

\end{titlepage}


\newpage
\par
\section{Introduction}

The purpose of this Letter is to justify the Abelian dominance in the low-energy region of QCD defined in the maximal Abelian (MA) gauge.  
This story begins with the idea of 't Hooft \cite{tHooft81} called the Abelian projection.  Immediately after this proposal, a hypothesis of Abelian dominance in low-energy physics of QCD was claimed by Ezawa and Iwazaki \cite{EI82}.
By adopting the MA gauge invented by Kronfeld et al. \cite{KLSW87}, actually, Abelian dominance was discovered by Suzuki and Yotsuyanagi \cite{SY90} a decade ago based on numerical simulation on a lattice and has been confirmed by the subsequent simulations, see \cite{review} for reviews.
However there is no analytical derivation or proof of Abelian dominance so far.
How can one justify or prove the Abelian dominance in low-energy physics in QCD?

\par
In a previous paper \cite{KondoI}, we tried to give an answer by constructing an effective Abelian gauge theory which is considered to be valid in the low-energy region of QCD.  We called it the Abelian-projected effective gauge theory (APEGT), although this name is somewhat misleading as will be explained below.
Before this work, a number of low-energy effective gauge theories were already proposed based on the idea of Abelian-projection.  However, we should keep in mind that these models were constructed by ignoring all the off-diagonal gluon fields from the beginning under the assumption of the Abelian dominance and/or the Abelian electro-magnetic duality, even if they can well describe some features of confinement physics in QCD.  In fact, they could not be derived by starting with the QCD Lagrangian.  Therefore, one can neither answer how the off-diagonal gluon fields influence the low-energy physics, nor how the Abelian electro-magnetic duality could appear from the non-Abelian gauge theory.
\par
In contrast to these models, the APEGT is a first-principle derivation of effective theory from QCD.   It was shown \cite{KondoI} that the off-diagonal gluons do affect the low-energy physics in the sense that off-diagonal gluons renormalize the resulting effective Abelian gauge theory.
  Moreover, the coupling constant of the effective Abelian gauge theory has the renormalization-scale dependence governed by the renormalization group $\beta$-function which is exactly the same as the original QCD, thereby, exhibiting the asymptotic freedom. In this sense, the APEGT reproduces a characteristic feature of the original QCD, asymptotic freedom, even if it is an Abelian gauge theory.
In addition, it was demonstrated how the dual Abelian gauge theory (magnetic theory) can in principle be obtained in the low-energy region of QCD.  Actually, it is possible to show \cite{KondoI} that monopole condensation leads to a dual Ginzburg-Landau theory supporting the dual superconductor picture \cite{Nambu74} of QCD vacuum.  
A version of the non-Abelian Stokes theorem indicates that the Wilson loop operator can be expressed in terms of diagonal gluon fields, see e.g. \cite{KT99}.
Combining these results, we are able to explain the Abelian dominance in quark confinement, see \cite{KT99}.  
\par
In the derivation of APEGT, however, we have treated the off-diagonal gluons as if they are massive in the MA gauge.  This assumption was necessary to justify the procedure of integrating out the off-diagonal gluon fields based on the functional integral, since this integration was interpreted as a step of the Wilsonian renormalization group (RG) of integrating out the massive (high-energy) modes.  In view of this, the resultant APEGT is regarded as the low-energy effective theory which is meaningful at least in the length scale $R>M_A^{-1}$ with $M_A$ being the mass of off-diagonal gluons.  In the derivation, moreover, we have integrated out the off-diagonal ghosts and anti-ghosts.  This step was also necessary to reproduce the correct coefficient of the $\beta$-function.
\par
To really justify the Abelian dominance, therefore, we need to show that the off-diagonal gluons and ghosts become massive in the MA gauge within the same framework as the APEGT.  
The main purpose of this Letter is to demonstrate that this is indeed the case.  
Actually, the derivation of off-diagonal gluon mass and ghost mass can be performed within the setting up of the previous paper \cite{KondoI}.
In the previous work, we have ignored the ghost self-interactions in the derivation of the APEGT, simply because they were not necessary to obtain the asymptotic freedom. 
In this Letter we properly take the ghost self-interaction into account.  We show that the quartic ghost self-interaction among off-diagonal ghosts leads to two kinds of ghost condensation. As a result, the off-diagonal gluons and off-diagonal ghosts (anti-ghosts) acquire non-zero masses.
\par
It should be remarked that the quartic ghost interaction term is generated by integrating out the off-diagonal gluon fields, even if such an interaction term is absent in the original Lagrangian of QCD.  This is due to the {\it nonlinearity} of the MA gauge.  In general, quartic ghost self-interaction terms are generated in the nonlinear gauge due to radiative corrections.  For the theory to be renormalizable, therefore, we need to incorporate quartic ghost self-interaction in the bare Lagrangian via the gauge-fixing and FP ghost term, as pointed out already in Appendix B of \cite{KondoI}. 
If so, how one can specify the quartic ghost interaction?  As a possibility, we introduce it so as to keep the {\it supersymmetry} \cite{KondoII} which is quite different from that of supersymmetric theory in theoretical particle physics.  It is hidden in the gauge fixing and ghost part of the MA gauge, while there is no supersymmetry in the Yang-Mills Lagrangian, since we are dealing with the usual QCD without supersymmetry.  This requirement determines almost uniquely the ghost self-interaction.
A special case was already examined in the previous paper \cite{KondoII,KondoIII,KondoIV,KondoV,KondoVI,KT99} in a slightly different context.
In this point, this Letter supplements the previous paper \cite{KondoI} by taking into account the ghost self-interactions properly.  
\par

\section{QCD in the modified MA gauge}

For the gauge group $G=SU(N)$, we consider the Cartan decomposition of the gauge potential into the diagonal and off-diagonal components,
\begin{equation}
  {\cal A}_\mu(x) = {\cal A}_\mu^A(x) T^A = a_\mu^i(x)T^i + A_\mu^a(x) T^a ,
\end{equation}
where $A=1,\cdots,N^2-1$.
Then the maximal Abelian (MA) gauge \cite{KLSW87} is defined as follows.  We define the functional of off-diagonal gluon fields,
$
  R[A] := \int d^4x {1 \over 2} A_\mu^a(x) A^\mu{}^a(x) .
$
The MA gauge is obtained by minimizing the functional $R[A^U]$ with respect to the local gauge transformation $U(x)$ of $A_\mu^a(x)$.
Then we obtain the differential form of the MA gauge,
\begin{equation}
  \partial_\mu A^\mu{}^a - g f^{abi} a_\mu^i A^\mu{}^b := D_\mu^{ab}[a]A^\mu{}^b = 0 .
\end{equation}
This is nothing but the the background-field gauge with the background field $a_\mu^i$.  After the MA gauge is adopted, the original gauge group $G=SU(N)$ is broken to the maximal torus group $H=U(1)^{N-1}$.
The MA gauge is a partial gauge fixing which fixes the gauge degrees of freedom for the coset space $G/H$.
\par
Following the well-known procedure, the manifest covariant action of QCD in the MA gauge is given by 
\begin{eqnarray}
  S_{QCD} &:=& S_{YM} + S_{GF+FP} + S_{F} ,
\nonumber\\
  S_{YM} &=& \int d^4x {-1 \over 4} {\cal F}_{\mu\nu}^A {\cal F}^{\mu\nu}{}^A ,
\nonumber\\
 S_{F} &=& \int d^4x \bar \Psi i \gamma^\mu (\partial_\mu - i g {\cal A}_\mu) \Psi ,
\nonumber\\
  S_{GF+FP} &=& - \int d^4x  i \delta_B [\bar C^a (D_\mu[a]A^\mu+{\alpha \over 2}B)^a ] .
\label{QCDL}
\end{eqnarray}
Here $S_{GF+FP}$  is the gauge fixing (GF) and Faddeev-Popov (FP) ghost term where $\delta_B$ is the Becchi-Rouet-Stora-Tyupin (BRST) transformation, $\alpha$ is the gauge fixing parameter and $B$ is the Nakanishi-Lautrup (NL) Lagrange multiplier field.  
Of course, we can add the gauge fixing term for the residual symmetry $H$ which we don't discuss in this Letter.

\par
In this paper, we adopt the {\it modified} MA gauge proposed in \cite{KondoII},
\begin{equation}
  S_{GF+FP}' = \int d^4x i \delta_B \bar \delta_B \left[ 
{1 \over 2} A_\mu^a(x) A^\mu{}^a(x) - {\alpha \over 2}i C^a(x) \bar C^a(x) \right] ,
\label{GF1}
\end{equation}
where $\delta_B$ ($\bar \delta_B$) is the BRST (anti-BRST) transformation.
The $\alpha=-2$ case has been already investigated in \cite{KondoII,KondoIV,KondoVI}.  The modified MA gauge is different from the naive MA gauge by the ghost self-interaction, since
\begin{equation}
  S_{GF+FP}' = - \int d^4x i \delta_B \left[ 
\bar C^a  \left\{ D_\mu[a]A^\mu + {\alpha \over 2} B \right\}^a
- i {\zeta \over 2} g f^{abi} \bar C^a \bar C^b  C^i
- i {\zeta \over 4} g f^{abc} C^a \bar C^b \bar C^c \right] ,
\label{GF2}
\end{equation}
where we must put $\zeta=\alpha$ to recover Eq.(\ref{GF1}).
Our choice of the Lagrangian (\ref{GF1}) (or (\ref{GF2}) with $\zeta=\alpha$) is invariant under the BRST and anti-BRST transformations.  Moreover, it is invariant for arbitrary $\alpha$ under the FP ghost conjugation (discrete symmetry) \cite{antiBRST},
\begin{equation}
 C^A \rightarrow \pm \bar C^A, \quad
 \bar C^A \rightarrow \mp C^A, \quad
 B^A \rightarrow - \bar B^A, \quad
 \bar B^A \rightarrow - B^A, \quad
 {\cal A}_\mu^A \rightarrow {\cal A}_\mu^A .
\end{equation}
Moreover, our choice (\ref{GF1}) leads to a renormalizable theory and preserves the hidden supersymmetry as discussed in the final part of this Letter.
The Lagrangian (\ref{GF2}) should be compared with the Lagrangian in the Lorentz gauge which is invariant under the FP conjugation only in the Landau gauge, 
\begin{equation}
 {\cal L}_{GF+FP} = -i \delta_B[\bar C^A(\partial_\mu {\cal A}^\mu+{\alpha \over 2}B)^A] = + i\delta_B \bar \delta_B({1 \over 2}{\cal A}_\mu^2)+{\alpha \over 2}i \bar \delta_B(B^A C^A) ,  
\end{equation}
although it is also invariant under the BRST and anti-BRST transformations.

\par
By performing the BRST transformation explicitly, we obtain
\begin{eqnarray}
  S_{GF+FP}' &=&   \int d^4x  \{ 
B^a D_\mu[a]^{ab}A^\mu{}^b+ {\alpha \over 2} B^a B^a
\nonumber\\
&&+ i \bar C^a D_\mu[a]^{ac} D^\mu[a]^{cb} C^b
- i g^2 f^{adi} f^{cbi} \bar C^a C^b A^\mu{}^c A_\mu^d 
\nonumber\\
&&+ i \bar C^a D_\mu[a]^{ac}(g f^{cdb}  A^\mu{}^d C^b)
+ i \bar C^a g  f^{abi} (D^\mu[a]^{bc}A_\mu^c) C^i 
\nonumber\\
&&+{\zeta \over 8} g^2 f^{abe}f^{cde} \bar C^a \bar C^b C^c C^d
+ {\zeta \over 4} g^2 f^{abc} f^{aid} \bar C^b \bar C^c C^i C^d
+ {\zeta \over 2} g f^{abc} i B^b C^a \bar C^c  
\nonumber\\
&&- \zeta  g f^{abi} i B^a \bar C^b C^i 
+ {\zeta \over 4} g^2 f^{abi} f^{cdi} \bar C^a \bar C^b C^c C^d \} ,
\label{GF3}
\end{eqnarray}
where the choice (\ref{GF1}) specifies the strength of the quartic ghost interactions,
\begin{equation}
  \zeta=\alpha .
\end{equation}
  An implication of this choice will be discussed later.  The gauge fixing term (\ref{GF3}) is the most general type we consider in the following.

\par
In particular, the $G=SU(2)$ case is greatly simplified as
\begin{eqnarray}
  S_{GF+FP}' &=&   \int d^4x  \{ 
B^a D_\mu[a]^{ab}A^\mu{}^b+ {\alpha \over 2} B^a B^a
\nonumber\\
&&+ i \bar C^a D_\mu[a]^{ac} D^\mu[a]^{cb} C^b
- i g^2 \epsilon^{ad} \epsilon^{cb} \bar C^a C^b A^\mu{}^c A_\mu^d 
\nonumber\\
&&
+ i \bar C^a g  \epsilon^{ab} (D_\mu[a]^{bc}A_\mu^c) C^3 
\nonumber\\
&&- \zeta  g \epsilon^{ab} i B^a \bar C^b C^3 
+ {\zeta \over 4} g^2 \epsilon^{ab} \epsilon^{cd} \bar C^a \bar C^b C^c C^d \} .
\label{GF4}
\end{eqnarray}
Integrating out the NL field $B^a$ leads to
\begin{eqnarray}
  S_{GF+FP}' &=&   \int d^4x  \{ 
-{1 \over 2\alpha}(D_\mu[a]^{ab}A^\mu{}^b)^2  
+ ( 1-\zeta/\alpha ) i \bar C^a g  \epsilon^{ab} (D^\mu[a]^{bc}A_\mu^c) C^3 
\nonumber\\
&&+i \bar C^a D_\mu[a]^{ac} D^\mu[a]^{cb} C^b
- i g^2 \epsilon^{ad} \epsilon^{cb} \bar C^a C^b A^\mu{}^c A_\mu^d 
\nonumber\\
&&+ {\zeta \over 4} g^2 \epsilon^{ab} \epsilon^{cd} \bar C^a \bar C^b C^c C^d \} .
\label{GF5}
\end{eqnarray}

\section{Ghost condensation and mass generation due to quartic ghost interaction}

\par
The  $\zeta=0$ case was considered in the previous paper \cite{KondoI}, leaving $\alpha$ arbitrary.  Even in this case, the quartic ghost self-interaction is generated after integrating out the off-diagonal gluons as mentioned above (see eq.(2.52) and Appendix B of \cite{KondoI}), 
\begin{eqnarray}
   z_{4c} g^2 \epsilon^{ab} \epsilon^{cd} \bar C^a \bar C^b C^c C^d,
\quad z_{4c} = 4 N {g^2 \over (4\pi)^2} \ln {\mu \over \mu_0}  \quad (N=2)  ,
\label{gsi}
\end{eqnarray}
since the interaction term 
$- i g^2 \epsilon^{ad} \epsilon^{cb} \bar C^a C^b A^\mu{}^c A_\mu^d$
does not vanish even for $\zeta=0$ (or $\alpha=0$).
\par
If we consider the non-zero $\zeta$ case, (\ref{gsi}) leads to the renormalization of $\zeta$ (together with $g$).  This is expected from the beginning, since the quartic ghost interaction is a renormalizable interaction.  In fact, it has been proven that QCD in the MA gauge is renormalizable by including the quartic ghost interaction\cite{MLP85}.

\par
For simplicity, we first discuss the $G=SU(2)$ case.\footnote{In the SU(2) case, the ghost condensation was seriously discussed by Schaden \cite{Schaden99} from a different viewpoint from ours.}
To incorporate the effect of ghost interaction, we introduce the auxiliary scalar field $\varphi$ as
\begin{eqnarray}
  {\zeta  \over 4} g^2 \epsilon^{ab} \epsilon^{cd} \bar C^a \bar C^b C^c C^d 
\rightarrow {-1 \over 2\zeta g^2} \varphi^2 -  \varphi i \epsilon^{ab} \bar C^a  C^b ,
\label{aux}
\end{eqnarray}
where we have used the identity,
$
   \epsilon^{ab} \epsilon^{cd} \bar C^a \bar C^b C^c C^d 
= 2 (i \epsilon^{ab} \bar C^a C^b)^2 =  2 (i \bar C^a C^a)^2 
$
$
= -4 \bar C^1 C^1 \bar C^2 C^2 .
$
Then the GF+FP term is cast into the form,
\begin{eqnarray}
  S_{GF+FP}' &=&   \int d^4x  \left\{ 
 i \bar C^a \partial_\mu \partial^\mu C^a
 -   \varphi i \epsilon^{ab} \bar C^a  C^b
- {1 \over 2\zeta g^2} \varphi^2  \right\} + \cdots .
\label{GF6}
\end{eqnarray}
\par
In order to see whether the QCD vacuum chooses a non-trivial $\varphi$ or not, we consider the effective potential for the $x$-independent $\varphi$ neglecting the kinetic terms.
The Coleman-Weinberg type argument (summing up all one-loop ghost diagrams with arbitrary number of external $\varphi$ fields)
\footnote{Note that the mathematical identity holds
$
 - {\rm tr} \sum_{n=1}^{\infty} {1 \over n}\left( 
{-i\varphi \over \partial^2} \right)^{2n} 
= \ln [1+ \varphi^2/(\partial^2)^2]
= \ln \det ( \partial_\mu \partial^\mu \delta^{ab} -    \varphi \epsilon^{ab} ) 
- \ln \det ( \partial_\mu \partial^\mu \delta^{ab}). 
$
}
 or integration over off-diagonal ghosts and anti-ghosts leads to the effective potential $V(\varphi)$ for $\varphi$,
\begin{eqnarray}
  V(\varphi) \int d^4x &=&   \int d^4x {1 \over 2\zeta g^2} \varphi^2  
+ i \ln \det ( \partial_\mu \partial^\mu \delta^{ab} -  \varphi \epsilon^{ab} ) .
\label{effpot}
\end{eqnarray}
Hence we obtain
\begin{eqnarray}
  V(\varphi)  &=&  {1 \over 2\zeta g^2} \varphi^2  
-   \int {d^4 k \over i(2\pi)^4} \ln [(-k^2)^2+ \varphi^2] .
\label{effpotE}
\end{eqnarray}
The stationary point is given by the zero of the gap equation,
\begin{eqnarray}
  V'(\varphi)   &\equiv&  \varphi \left[   {1 \over \zeta g^2}   
- 2 \int {d^4 k  \over i(2\pi)^4} {1 \over (-k^2)^2+\varphi^2} \right] 
= 0 .
\label{effpotEd}
\end{eqnarray}
Within the minimal subtraction (MS) scheme of the dimensional regularization, the effective potential is obtained as%
\begin{equation}
 V(\varphi) = {1 \over 2\zeta g^2} \varphi^2 + {1 \over 32\pi^2}   \varphi^2 \left[ 2 \ln \left({1 \over 4\pi \mu^2} |\varphi| \right)  + C  \right] ,
\label{poten}
\end{equation}
with  $C := 2 \gamma - 3$ and the Euler constant $\gamma=0.5772\cdots$.
As far as $\zeta \not= 0$,   
the gap equation (\ref{effpotEd}) has   non-trivial solutions given by 
$\varphi=\pm \varphi_0$ (besides a trivial one $\varphi=0$) where
\begin{equation}
  v :=   \varphi_0 = 4\pi \mu^2 e^{1-\gamma}\exp \left[ {-8\pi^2 \over \zeta  g^2(\mu)} \right]   > 0 .
\end{equation}
These solutions correspond  to global minima of the effective potential. 
At the global minimum $\varphi=\varphi_0$, $V(\varphi_0)=-{1 \over 32\pi^2} (\varphi_0)^2 < 0$.
This shows that QCD vacuum prefers a ghost condensate for any value of $\zeta  g^2(\not=0)$ such that
$
  \varphi_0 \sim \zeta  g^2  \langle i \epsilon^{ab} \bar C^a  C^b \rangle  \not= 0 .
$
\par
Now we consider the theory around the nontrivial vacuum $\varphi=\varphi_0$. 
The off-diagonal ghost propagator is modified in the ghost-condensed vacuum into
\begin{equation}
  \langle C^a(x)  \bar C^b(y) \rangle  
=  i \int {d^4k \over i(2\pi)^4} {-k^2 \delta^{ab}-v \epsilon^{ab} \over (-k^2)^2+v^2} e^{i k(x-y)} .
\end{equation}
When $\langle \epsilon^{ab} \bar C^a  C^b \rangle=0$, i.e., 
$v=0$,
another ghost condensation 
$\langle \bar C^a  C^a \rangle$ is also zero in the dimensional regularization. 
However, non-zero condensation 
$\langle \epsilon^{ab} \bar C^a  C^b \rangle  \not= 0$
leads to another condensation 
$\langle \bar C^a  C^a \rangle \not= 0$.
In the condensed vacuum, the ghost-gluon 4-body interaction, 
$
    -i g^2 \epsilon^{ad} \epsilon^{cb} \bar C^a C^b A^\mu{}^c A_\mu^d ,
$
leads to a mass term of the off-diagonal gluons, 
\begin{equation}
    -i g^2 \epsilon^{ad} \epsilon^{cb} 
\langle \bar C^a C^b \rangle A^\mu{}^c A_\mu^d  
=   {1 \over 2} g^2 \langle i \bar C^c C^c \rangle A^\mu{}^a A_\mu^a ,
\end{equation}
where we have used 
$
  \langle \bar C^a  C^b \rangle 
= {1 \over 2}(\delta^{ab} \langle \bar C^c  C^c \rangle
+ \epsilon^{ab} \langle \epsilon^{cd} \bar C^c  C^d \rangle ) .
$ 
The condensation is given by
\begin{eqnarray}
  \langle i \bar C^a C^a \rangle  
=    \int {d^4k \over i(2\pi)^4} {-2k^2 \over (-k^2)^2+v^2}  
=  {v \over 16\pi}  > 0 ,
\end{eqnarray}
where the signature, i.e., positivity of $v$ is determined by analytic  continuation  to Euclidean region.
Thus the off-diagonal gluon acquires the mass given by
\begin{eqnarray}
   M_A^2 =   g^2 \langle i \bar C^a C^a \rangle  
= {g^2v \over 16\pi} > 0 .
\label{gluonmass}
\end{eqnarray}
The dynamically generated mass $m_A$ is finite excluding the mass counter term.
Note that the introduction of the explicit mass term ${1 \over 2}m^2 A_\mu^a A^\mu{}^a$ spoils the renormalizability of the 
theory.\footnote{Even if one introduces a bare mass term of the form,
$
  m^2 \left( {1 \over 2}A_\mu^a A^\mu{}^a + i\alpha \bar C^a C^a \right)
$,
the modified BRST and anti-BRST transformations can be constructed as
$
 \delta_B B = m^2 C,
 \bar \delta_B B =  m^2 \bar C - g(\bar C \times B) ,
$
under which the modified Lagrangian is invariant.
However, the nilpotency of both transformations is violated as
$
 \delta_B^2 \bar C = im^2 C,
\bar \delta_B^2 C = -im^2 \bar C ,
$
leading to the breakdown of physical $S$-matrix unitarity, see \cite{Ojima82}.
}
Our derivation of off-diagonal gluon mass preserves the renormalizability, see \cite{KS00} for more details.
\par
Now we proceed to estimate the order of the off-diagonal mass.
We impose the renormalization condition at the renormalization point $M$, i.e, we define the renormalized coupling $\zeta g^2$ by
\begin{equation}
 V''(\varphi)|_{\varphi=M^2} = (\zeta g^2)^{-1}_{(M)} .
\end{equation}
where $M$ is nonzero but arbitrary.
The renormalizability implies that arbitrary choice of $M$ should not change the physics.  Hence we have the $\mu$-independence of $V(\varphi)$ which means that
$(\zeta g^2)^{-1}_{(M)}-(4\pi^2)^{-1} \ln M$ is a $M$-independent constant.  Then we have
$(\zeta g^2)_{(M)}=(\zeta g^2)_{(M_0)}[1+(\zeta g^2)_{(M_0)} (4\pi^2)^{-1} \ln {M \over M_0}]^{-1}$.  Hence $\zeta g^2$ satisfies the RG equation,
$
 M {d \over dM}(\zeta g^2) = - {1 \over 4\pi^2}(\zeta g^2)^2 .
$
From the asymptotic freedom, 
$
 M {d g^2 \over dM} = - {b_0 \over 8\pi^2}g^4 ,
$
$\zeta$ obeys the RG equation,
$
 M {d \zeta \over dM} = - {g^2 \over 4\pi^2}\zeta (\zeta-b_0/2) .
$
Then we find that $\zeta=0$ and $\zeta=b_0/2$ are the fixed points.
The $M$ independence of  $V$ means that $V$ satisfies the differential equation,
\begin{equation}
  \left[ M{\partial \over \partial M} + \beta(g){\partial \over \partial g}
+ \gamma_\zeta(g){\partial \over \partial \zeta}
- \gamma_\varphi(g) \varphi {\partial \over \partial \varphi} \right] V(\varphi) = 0 ,
\label{RGeq}
\end{equation}
where $\gamma_\varphi(g)$ is the anomalous dimension of $\varphi$ defined by
$\gamma_\varphi(g):={1 \over 2}M{\partial \ln Z_\varphi \over \partial M}$ and $\varphi_R :=Z_\varphi^{-1/2}\varphi$
and 
$\gamma_\zeta(g):=M{\partial \zeta \over \partial M}$.
  Substituting  (\ref{poten}) into (\ref{RGeq}), we obtain the consistent results,
$
 \gamma_\varphi(g)=-{b_0 g^2 \over 16\pi^2} ,
$
$
 \beta(g) = g \gamma_\varphi(g) = -{b_0 g^3 \over 16\pi^2},
$
and
$
 \gamma_\zeta(g):= - {g^2 \over 4\pi^2}\zeta (\zeta-b_0/2).
$
The non-trivial fixed point $\zeta=b_0/2$ yields 
$
  v  = 4\pi e^{1-\gamma} \mu^2  \exp \left( -{16\pi^2 \over b_0  g^2(\mu)} \right) 
$
$
 = 4\pi e^{1-\gamma} \Lambda_{QCD}^2 .
$
Therefore, the condensation $v$ is a renormalization-group invariant and the order is given by the QCD scale, $\Lambda_{QCD}$.
Hence, the off-diagonal gluon mass is given by
$
 M_A={g \over 2}e^{(1-\gamma)/2} \Lambda_{QCD}
= (\pi \alpha_s)^{1/2} e^{(1-\gamma)/2} \Lambda_{QCD}.
$
This is comparable with the Lattice simulation result, $M_A \cong 1.2 {\rm GeV}$, see \cite{AS99}.  
\par
Moreover, the quartic ghost interaction can give a mass for the ghost, 
since the treatment $\grave{a}$ la Hartree-Fock approximation leads to
$
  {\zeta  \over 4} g^2 \epsilon^{ab} \epsilon^{cd} \bar C^a \bar C^b C^c C^d 
=   {\zeta  \over 2} g^2 (i \epsilon^{ab} \bar C^a C^b)^2
=  {\zeta  \over 2} g^2 (i \bar C^a C^a)^2
\rightarrow    \zeta  g^2 \langle i \bar C^a C^a \rangle
i \bar C^b C^b .
$
This implies the off-diagonal ghost mass,
\begin{eqnarray}
   M_c^2 \cong  \zeta   g^2 \langle i \bar C^a C^a \rangle  = \zeta g^2 {v \over 16\pi}  = \zeta M_A^2 .
\label{ghostmass}
\end{eqnarray}
Thus off-diagonal gluons and ghosts can become massive due to ghost self-interactions. 
Note that $\bar C^a C^a$ and $\epsilon^{ab} \bar C^a  C^b$ are invariant under the residual U(1).  Even in the presence of the condensation, the residual U(1) invariance is not broken spontaneously and the diagonal gluon remains massless \cite{KS00}.
 These results strongly support  the Abelian dominance.

\par
It is possible to extend the above analysis to the SU(3) case \cite{KS00}.  The potential $V(\varphi^3,\varphi^8)$ is written in terms of two diagonal combinations, $\varphi^i \sim \zeta g^2 \sqrt{-1} f^{iab}\bar C^a C^b(i=3,8)$. 
In fact, the effective potential for $SU(3)$ is given by
\begin{eqnarray}
  V(\vec \varphi)  &=&  {1 \over 2\zeta g^2} \vec \varphi \cdot \vec \varphi    
-   \sum_{\alpha=1}^{3} \int {d^4 k \over i(2\pi)^4} \ln [(-k^2)^2+ 
(\vec \epsilon_\alpha \cdot \vec \varphi)^2] ,
\label{effpotEsu3}
\end{eqnarray}
where $\vec \varphi:=(\varphi^3,\varphi^8)$ and $\epsilon_\alpha$ is the root vectors 
$
\epsilon_1=(1,0), \epsilon_2 = (-{1 \over 2},-{\sqrt{3} \over 2}), \epsilon_3 = (-{1 \over 2},{\sqrt{3} \over 2}).
$
The schematic plot of the potential is given in Fig.1. 
It turns out that the potential has the global minima at six points on  the three straight lines along the root vectors, i.e., 
(I) $\varphi^8 = 0, \varphi^3\not=0$, 
(II) $\varphi^8= \sqrt{3}\varphi^3$,
(III) $\varphi^8= -\sqrt{3}\varphi^3$.
We find that the off-diagonal gluons in the $SU(3)$ case have two different masses as follows.\footnote{This result is obtained up to Weyl symmetry.}
\begin{eqnarray}
{\rm I:} &&  {1 \over \sqrt{2}}m_{A^1}={1 \over \sqrt{2}}m_{A^2}= m_{A^4}=m_{A^5}=m_{A^6}=m_{A^7},
\nonumber\\
{\rm II:} &&   m_{A^1} = m_{A^2}  ={1 \over \sqrt{2}} m_{A^4} ={1 \over \sqrt{2}} m_{A^5} = m_{A^6} = m_{A^7} ,
\nonumber\\
{\rm III:} &&  m_{A^1} = m_{A^2}  = m_{A^4} = m_{A^5} ={1 \over \sqrt{2}} m_{A^6} ={1 \over \sqrt{2}} m_{A^7} .
\end{eqnarray}
The value of larger mass is given by
$
 m_A^2 = {g^2 V_0 \over 16\pi}, 
$
where
$
V_0 = 4^{1/6} (4\pi \mu^2) e^{1-\gamma} \exp \left(-{16\pi^2 \over 3\zeta g^2}\right) .
$
The $\mu$-independence of the potential holds when $\zeta=b_0/3$ for $SU(3)$.  Hence, we obtain
$
V_0 = 4^{1/6} (4\pi ) e^{1-\gamma} \Lambda_{QCD}^2 .
$
Another way to estimate the order of the off-diagonal mass is based on the identity of the trace anomaly,
\begin{equation}
  \langle 0|  T^\mu_\mu |0 \rangle 
= {\beta(\alpha_s) \over 4\alpha_s } \langle 0| ({\cal F}^A_{\mu\nu})^2 |0 \rangle
= - {11N_c -2N_f \over 24}
\langle 0| {\alpha_s \over \pi}({\cal F}^A_{\mu\nu})^2 |0 \rangle .
\label{tra}
\end{equation}
Note that the values of gluon condensate obtained on a lattice are as follows.\footnote{
The authors would like to thank E.-M. Ilgenfritz for providing this information. 
In the presence of light quarks, the charmonium sum rules \cite{FPS84} gives
$
\langle 0| {\alpha_s \over \pi}({\cal F}^A_{\mu\nu})^2 |0 \rangle
\cong 1.3 \sim 1.9 \times 10^{-2} {\rm (GeV)^4} .
$ 
}
$
\langle 0| {\alpha_s \over \pi}({\cal F}^A_{\mu\nu})^2 |0 \rangle 
 \sim 0.152  {\rm (GeV)^4} 
$ 
for $G=SU(2)$ and 
$
0.144  {\rm (GeV)^4} 
$ 
for $G=SU(3)$.
On the other hand, the vacuum energy of the condensed vacuum (at the global minima) leads to
\begin{equation}
 \langle 0|  T^\mu_\mu |0 \rangle \cong 4V(\varphi_0) = - {3V_0^2 \over 16\pi^2} \mbox{for SU(3)},
\left(-  {v^2 \over 8\pi^2} \mbox{for SU(2)} \right).
\label{vae}
\end{equation}
Equating (\ref{tra}) and (\ref{vae}), we obtain
$
 V_0 = (|\langle 0|  T^\mu_\mu |0 \rangle|16\pi^2/3)^{1/2} \cong 3.2 {\rm (GeV)^2} ,
$
for $N_c=3$ and $N_f=0$.  Finally, we have 
$
 m_A = (\alpha_s V_0/4)^{1/2},  (\alpha_s V_0/8)^{1/2} \cong 0.4 \sim 0.5  {\rm GeV} .
$
These are our predictions.
The full details of $SU(3)$ case will be given in \cite{KS00}.

\par
\begin{figure}
\begin{center}
 \leavevmode
 \epsfxsize=60mm
 \epsfysize=60mm
 \epsfbox{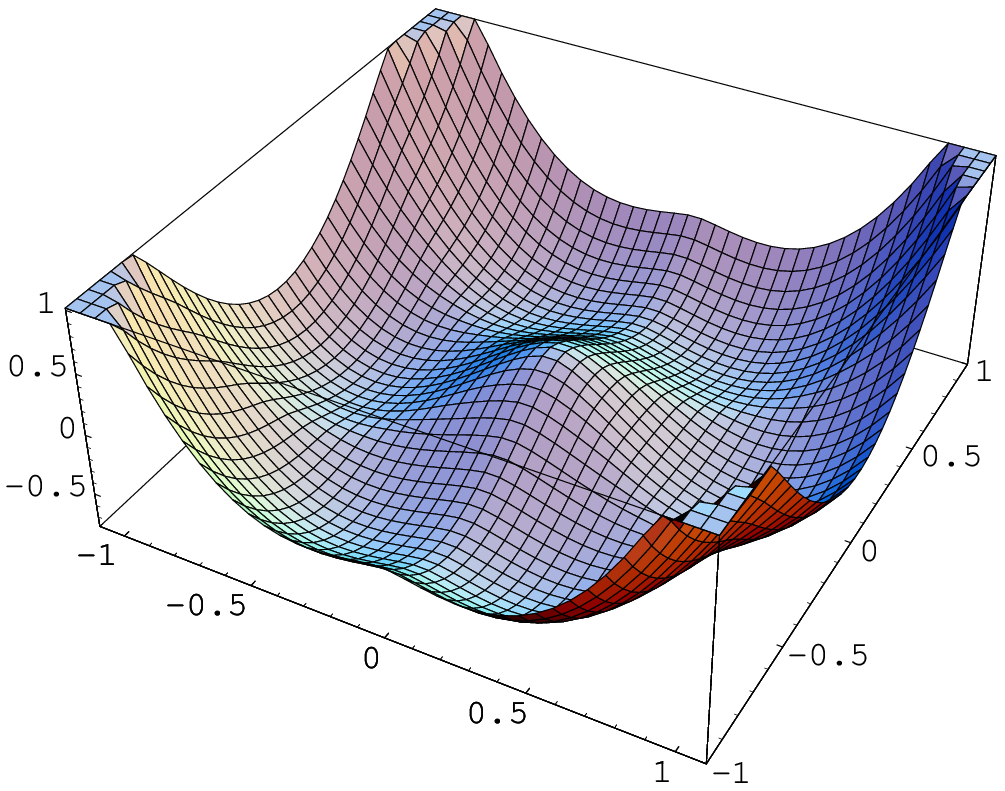}
 \epsfxsize=60mm
 \epsfysize=60mm
 \epsfbox{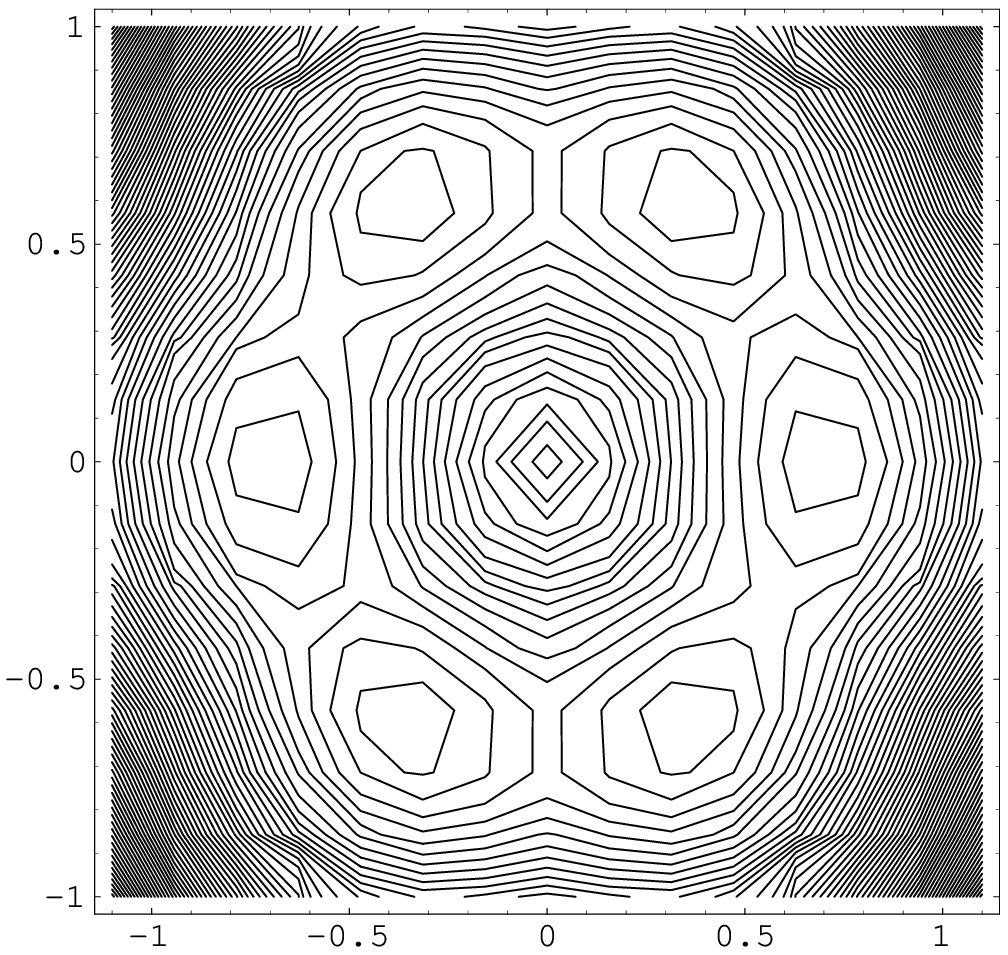}
\end{center} 
 \caption[]{The three-dimensional plot of the effective potential $V(\varphi^3,\varphi^8)$ and its contour plot.}
 \label{Fig.1}
\end{figure}

\par
\section{APEGT of QCD in the modified MA gauge}
\par
In order to obtain the ``effective" theory which is written in terms of the diagonal fields $a^i_\mu, B^i, C^i, \bar C^i$ alone, we intend to integrate out all the off-diagonal fields $A^a_\mu, B^a, C^a, \bar C^a$.  We call the resultant effective field theory  the Abelian-projected effective gauge theory (APEGT).
That is to say, the APEGT is defined as
\begin{equation}
  \exp (i S_{APEGT}) 
= \int [dA_\mu^a][dC^a][d\bar C^a][dB^a] \exp (i S_{QCD}) .
\end{equation}
Hence the vacuum-to-vacuum amplitude (or the partition function) of QCD reads
\begin{eqnarray}
  Z_{QCD} &:=& \int [d{\cal A}_\mu^A][dC^A][d\bar C^A][dB^A] \exp (i S_{QCD}) 
\nonumber\\
&=& \int [da_\mu^i][dC^i][d\bar C^i][dB^i] \exp (i S_{APEGT}) .
\end{eqnarray}
 In the naive MA gauge, such an attempt was first performed by Quandt and Reinhardt \cite{QR98} for $\alpha=0$ and subsequently by one of the authors \cite{KondoI} for $\alpha \not= 0$, in particular, $\alpha=1$ \cite{KondoI} at least for $G=SU(2)$.  
(We have found that the $\alpha=0$ case is very special from the viewpoint of renormalizability.)
The generalization to $SU(N)$ is straightforward \cite{KS98,KS00}.
\par
In the naive MA gauge \cite{KondoI}, the off-diagonal gluons were expected to become massive, while the diagonal gluons were believed to behave  in rather complicated way.  Recently, the massiveness of off-diagonal gluons has been shown by Monte Carlo simulations on a lattice \cite{AS99}.  An analytical explanation was given at least in the topological sector based on the dimensional reduction of the topological sector to the two-dimensional coset $G/H$ nonlinear sigma (NLS) model, see section IV. C of \cite{KondoII}.  In this paper we have given another evidence of mass generation of off-diagonal gluons and ghosts.
In view of these facts, the integration of massive off-diagonal gluon fields can be interpreted as a step of integration of massive modes in the sense of the Wilsonian renormalization group.
In this sense, the APEGT obtained in this way is regarded as the low-energy effective theory describing the physics in the length scale $R>m_A^{-1}$ or in the low-energy region $p<m_A$.

\par
In order to obtain the explicit form of the APEGT in the modified MA gauge, we repeat the steps performed in \cite{KondoI} to obtain the APEGT. The GF+FP term in the condensed vacuum reads
\begin{eqnarray}
  S_{GF+FP}' &=&   \int d^4x  \Big[
-{1 \over 2\alpha}(D_\mu[a]^{ab}A^\mu{}^b)^2  
+ {1 \over 2}M_A^2 A^\mu{}^a A_\mu^a 
\nonumber\\
&+&i \bar C^a D_\mu[a]^{ac} D^\mu[a]^{cb} C^b
 - g \varphi_0 i \epsilon^{ab} \bar C^a C^b 
-g \tilde \varphi i \epsilon^{ab} \bar C^a C^b 
 - V(\varphi_0+\tilde \varphi) \Big] ,
\label{GF8}
\end{eqnarray} 
where we have put $\varphi=\varphi_0+\tilde \varphi$.  Note that 
$
 V(\varphi_0+\tilde \varphi) = V(\varphi_0) + {1 \over 2}\tilde \varphi^2 V''(\varphi_0) + O(\tilde \varphi^3)
$
with $V'(\varphi_0)=0$ and $V''(\varphi_0)={1 \over 8\pi^2}$.
We perform the integration over (high-energy) massive modes, i.e., off-diagonal gluons $A_\mu^a$ and off-diagonal ghosts $C^a$ and anti-ghosts $\bar C^a$ for the total action $S_{YM}+S_{GF+FP}'$.
In the process of deriving the APEGT, we have introduced the anti-symmetric auxiliary (Abelian) tensor field $B_{\mu\nu}^i$ to avoid  the quartic self-interactions among the off-diagonal gluons appearing in $S_{YM}$ where $B_{\mu\nu}^i$ is invariant under the residual gauge transformation $H=U(1)^{N-1}$.  The way of introducing $B_{\mu\nu}^i$ is not unique, see \cite{KondoI} and \cite{KS00} for more details.  In the following we discuss one of the original versions \cite{KondoI}.
By repeating the procedures in \cite{KS00}, we can show that the resultant APEGT is written as (up to higher-derivative terms)\footnote{The higher-derivative terms are suppressed in the low-energy region, since they are of the order  $O(p^2/M_A^2)$.
}
\begin{eqnarray}
  {\cal L}_{aB}[a,B] &=& -{1 \over 4g^2(\mu)} f_{\mu\nu}^i f^{\mu\nu}{}^i  
-{1+z_b \over 4} g^2 B_{\mu\nu}^i B^{\mu\nu}{}^i 
+ {z_c \over 2}  B_{\mu\nu}^i {}^*\! f^{\mu\nu}{}^i ,
\label{LaB}
\end{eqnarray}
where we have defined 
$
  g(\mu) := Z_a^{1/2} g 
$
with
$
Z_a := 1-z_a+z_d= 1+ {22 \over 3}N {g^2 \over (4\pi)^2} \ln {\mu \over \mu_0} .
$
Here $f_{\mu\nu}^i$ is the Abelian field strength 
$f_{\mu\nu}^i := \partial_\mu a_\nu^i - \partial_\nu a_\mu^i$ and ${}^*\!f_{\mu\nu}^i$ is the Hodge dual of $f_{\mu\nu}^i$, i.e.,
${}^*\!f_{\mu\nu}^i := {i \over 2} \epsilon_{\mu\nu\rho\sigma}f^{\rho\sigma}{}^i$.
This result shows that the off-diagonal gluons can not be ignored and that they influence the APEGT in the form of renormalization of the Abelian sector.  In fact, the renormalization factors $z_a, z_b, z_c, z_d$ are given by
$
 z_a = - {20 \over 3}N {g^2 \over (4\pi)^2} \ln {\mu \over \mu_0} , \quad
 z_b = + 2N {g^2 \over (4\pi)^2} \ln {\mu \over \mu_0} , \quad
 z_c = + 4N {g^2 \over (4\pi)^2} \ln {\mu \over \mu_0} , \quad
 z_d={2 \over 3}N {g^2 \over (4\pi)^2} \ln {\mu \over \mu_0}
$
where $\mu$ is a renormalization scale.
\par

\par
A remarkable fact is that the running of the gauge coupling constant $g(\mu)$ is governed by the  $\beta$-function,
\begin{equation}
  \beta(g) := \mu {dg(\mu) \over d\mu} = - b_0 g^3(\mu) ,
\quad b_0 = {11 \over 3}N > 0 ,
\label{beta}
\end{equation}
which is the same as the original Yang-Mills theory.
So the APEGT is an effective {\it Abelian} gauge theory exhibiting the asymptotic freedom.
The coupling between $B_{\mu\nu}^i$ and ${}^*\! f_{\mu\nu}{}^i$ is important to derive the dual Abelian gauge theory which leads to the dual superconductivity. 
This term is generated through the integration (or radiative corrections) and is absent in the original Lagrangian.  In this sense, the APEGT just obtained is non-renormalizable.  Nevertheless, the APEGT can be made renormalizable, see \cite{KS00} for more details.  
The effect of dynamical quarks can be included into this scheme by integrating out the quark fields.  It results in further renormalization leading to the $\beta$-function with a different coefficient, $b_0={11 \over 3}N - {4 \over 3}f r_F$, where $f$ is the number of quark flavors and $r_F$ is the dimension of fermion representation.
\par

\section{New extended BRS algebra}

It is easy to show that the QCD Lagrangian (\ref{QCDL}) in the modified MA gauge (\ref{GF4}) or (\ref{GF5}) has a new global symmetry if it is restricted to $C^3=0$ subspace or to the parameter $\zeta=\alpha$, that is to say, the Lagrangian is invariant under the two transformations, 
\begin{eqnarray}
  \delta_{+} \bar C^a(x) = C^a(x), \quad
 \delta_{+}({\rm other~fields}) = 0 ,
\\
  \delta_{-} C^a(x) =  \bar C^a(x), \quad
 \delta_{-}({\rm other~fields}) = 0 .
\end{eqnarray}
The existence of this symmetry in the Lagrangian in the maximal Abelian gauge was recently noticed by Schaden \cite{Schaden99}.
After eliminating $B^a$ (and putting $\zeta=\alpha$), (\ref{GF5}) agrees with the Lagrangian  examined by Schaden \cite{Schaden99} from a quite different viewpoint, the equivariant cohomology \cite{Schaden98}.
These transformations $\delta_{\pm}$ for the field $\Phi$ are defined by the generators $Q_{\pm}$ as 
$
 \delta_{\pm} \Phi = [iQ_{\pm}, \Phi] , \quad Q_{\pm} := \int d^3x J_{\pm}^0 ,
$
where the generators are constructed through the Noether currents,
\begin{eqnarray}
  J_{+}^\mu &=& - i C^a (D_\mu[a]C)^a = + i \delta_B(C^a A_\mu^a) ,
\nonumber\\
  J_{-}^\mu &=& + i \bar C^a (D_\mu[a]\bar C)^a = - i \bar \delta_B(\bar C^a A_\mu^a) .
\end{eqnarray}
They should be compared with the ghost number,
\begin{eqnarray}
  \delta_{c} C^A(x) &:=& [iQ_c, C^A(x)] =   C^A(x),
\nonumber\\
  \delta_{c} \bar C^A(x)  &:=& [iQ_c, \bar C^A(x)] =  - \bar C^A(x),
\end{eqnarray}
where $Q_c$ is the ghost charge defined by
$
 Q_c = \int d^3x J_c^0, \quad 
J_c^\mu = i \{ -(D^\mu[a]\bar C)^A C^A + \bar C^A (D^\mu[a]C)^A  \}  .
$
Shaden found that there is a SL(2,R) symmetry among $Q_{+}, Q_{-}$ and $Q_c$, i.e.,
$
 [iQ_c, Q_{+}] = 2Q_+, \quad
 [iQ_c, Q_{-}] = - 2Q_{-}, \quad 
i[Q_{+}, Q_{-}] = Q_c ,
$
where the diagonal generator is the ghost number $Q_c$.
\par
It is well known that the BRST transformation, anti-BRST transformation and the ghost number generator form the double BRS  algebra among three generators, $Q_B, \bar Q_B$ and $Q_c$,
\begin{eqnarray}
 && [Q_c, Q_c] = 0, 
\nonumber\\
 \{ Q_B, Q_B \} &=& 0, \quad \{ \bar Q_B, \bar Q_B \} = 0,
\nonumber\\
i[Q_c, Q_B] &=& Q_B, \quad i[Q_c, \bar Q_B] = - \bar Q_B ,
\nonumber\\
 && \{ Q_B, \bar Q_B \} = 0 .
\end{eqnarray}
By enlarging the double BRS  algebra, we find a new extended double BRS  algebra \cite{Kondo00} among five generators, $Q_B, \bar Q_B, Q_+, Q_-$ and $Q_c$, supplemented by
\begin{eqnarray}
 [Q_B, Q_+] &=& 0, \quad i[\bar Q_B, Q_+] = - Q_B ,
\nonumber\\
i[Q_B, Q_-] &=& - \bar Q_B, \quad  [\bar Q_B, Q_-] = 0 ,
\\
 i[Q_c, Q_+] &=& 2 Q_+, \quad i[Q_c, Q_-] = -2 Q_-,
\nonumber\\
&& i[Q_+, Q_-] = Q_c  .
\end{eqnarray}
Note that the new extended BRS algebra closes only on the space of functionals which are invariant under the residual U(1) gauge transformation.
\par

This should be compared with the extended BRS  algebra (BRSNO algebra) found by Nakanishi and Ojima \cite{NO80} in the manifest covariant gauge of the Lorentz type where the additional symmetry is given by
\begin{eqnarray}
\delta_{cc} B^A = -ig (C \times C)^A, 
\quad \delta_{cc} \bar C^A = -2 C^A, 
\quad \delta_{cc} ({\rm other~fields}) = 0, 
\nonumber\\
\delta_{\bar c \bar c} B^A = +ig (\bar C \times \bar C)^A, 
\quad \delta_{\bar c \bar c} C^A = +2 \bar C^A, 
\quad \delta_{\bar c \bar c} ({\rm other~fields}) = 0 . 
\end{eqnarray}
Although the BRSNO algebra holds for arbitrary gauge, their generators are conserved only in the Landau gauge $\alpha=0$.
In the new extended algebra given above, the generators are conserved for an arbitrary gauge parameter $\alpha$, but only on the space which is invariant under the residual gauge group. 

\section{Spontaneous breaking of a global symmetry and hidden supersymmetry in MA gauge}
\par
The non-zero expectation value 
$\langle  \epsilon^{ab} C^a \bar C^b \rangle$ is regarded as the spontaneous breaking of the SL(2,R) symmetry as pointed out by Schaden \cite{Schaden99}, since
\footnote{We can consider other types of ghost condensations with the non-zero ghost number, i.e.,
$
 \langle 0|[iQ_+, \epsilon^{ab}  C^a \bar C^b] |0 \rangle
=  \langle 0| \epsilon^{ab} C^a  C^b |0 \rangle
$
and
$   
  \langle 0|[iQ_-, \epsilon^{ab}  C^a \bar C^b] |0 \rangle
=  \langle 0| \epsilon^{ab} \bar C^a  \bar C^b |0 \rangle .   
$
Three composite operators $\epsilon^{ab} C^a \bar C^b, \epsilon^{ab} C^a  C^b, \epsilon^{ab} \bar C^a  \bar C^b$ are mutually related by the action of the generators $Q_+$ or $Q_-$ which are spontaneously broken.
}
\begin{equation}
 \langle 0|[iQ_+, \epsilon^{ab} \bar C^a \bar C^b] |0 \rangle
= 2 \langle 0| \epsilon^{ab} C^a \bar C^b |0 \rangle  
= - \langle 0|[iQ_-, \epsilon^{ab} C^a C^b] |0 \rangle .
\end{equation}
The non-compact SL(2,R) symmetry is spontaneously broken into the non-compact Abelian subgroup, since the ghost charge $Q_c$ is not broken.  
The massless Nambu-Goldstone (NG) particles associated with this spontaneous symmetry breaking can be confined by the quartet mechanism \cite{KO79}, i.e., decouple from physical observables, since the current $J^\mu_+$ ($J^\mu_-$) is BRST (anti-BRST) exact.  Therefore, we need not to worry about the emergence of massless particles.
\par
In the previous paper we have argued that the non-zero mass for the off-diagonal gluons can be understood from the massive spectrum of the coset NLS model in two dimensions, since  the GF and FP ghost part for the modified MA gauge in four dimensions is reduced to the coset NLS model in two dimensions by the dimensional reduction $\grave{a}$ la Parisi and Sourlas \cite{PS79}.
In this Letter we have argued that the quartic ghost interaction is an origin of off-diagonal gluon mass.  Now we discuss how two pictures could be related to each other.  
\par
It is shown that the action (\ref{GF1}) for gauge fixing and FP ghost in the modified MA gauge has the orthosymplectic symmetry $OSp(4|2)$ among $A_\mu^a, C^a, \bar C^a$ when it is written in the superspace $X^M := (x_\mu, \theta, \bar \theta)$ following the superspace formulation by Bonora and Tonin \cite{BT81}.
This superspace formulation can give a geometric meaning of BRST $\delta_B$ and anti-BRST $\bar \delta_B$ transformations as translations in the Grassmann variables  $\theta$ and $\bar \theta$ respectively,
$
  \delta_B \leftrightarrow {d \over d\theta} \leftrightarrow \int d\theta,
\quad 
  \bar \delta_B \leftrightarrow {d \over d\bar \theta} \leftrightarrow \int d 
\bar \theta ,
$
where we have employed the equivalence between the differentiation and integration with respect to the Grassmann variable.
Then the GF and FP part in the modified MA gauge is rewritten into the manifest $OSp(4|2)$ invariant form,
\begin{equation}
  S_{GF+FP}' = i \int d^4x \int d\theta \int d\bar \theta \
{\rm tr}_{G/H} \left[ {1 \over 2} \eta_{NM} {\cal A}^N(x,\theta,\bar \theta) {\cal A}^M(x,\theta,\bar \theta) \right] ,
\label{ssf}
\end{equation}
using the Lie-algebra valued superfield (one-form),
\begin{equation}
  {\cal A}_M (X) dX^M = {\cal A}_\mu(x,\theta,\bar \theta) dx^\mu
+ C(x,\theta,\bar \theta) d\theta + \bar C(x,\theta,\bar \theta) d\bar \theta ,
\end{equation}
and a supermetric
$ \eta_{NM} = \delta_{\mu\nu}$ for $(M,N)=(\mu,\nu)$
and 
$-i {\alpha \over 2}$ for $(M,N)=(\theta,\bar\theta)$.
Thanks to the $OSp(4|2)$ invariance of the integrand, it is shown \cite{KondoII} that (\ref{ssf}) is reduced to
\begin{equation}
  S_{GF+FP}' = \pi \alpha \int d^2z  \left[ {1 \over 2}  A_\mu^a(z) A^\mu{}^a(z) 
- {\alpha \over 2}i C^a(z) \bar C^a(z) \right] .
\label{reduced}
\end{equation}
If we restrict the gauge potential to its topological nontrivial piece in the coset $G/H$,
$
  A_\mu^a \rightarrow {\rm tr} \left[ T^a {i \over g} U(x) \partial_\mu U(x)^\dagger \right] := {1 \over g} \Omega_\mu^a ,
$
the action (\ref{reduced}) is nothing but the coset NLS model  \cite{KondoII,KT99},
\begin{equation}
  S_{GF+FP} = {\pi \alpha \over 2g^2} \int d^2z   \Omega_\mu^a(z) \Omega^\mu{}^a(z) .
\label{NLS}
\end{equation}
Thus, as far as $\alpha\not= 0$, the dimensional reduction to the two-dimensional coset NLS model occurs and the massive spectrum in the coset NLS model implies the massive off-diagonal gluon, see section IV.C of 
\cite{KondoII}.\footnote{In the $\alpha=0$ case, the $OSp(4|2)$ invariance is lost and hence the above mechanism of dimensional reduction does not work.  On the other hand, the quartic ghost interaction disappears in this case and the ghost condensation generating the off-diagonal gluon mass does not occur and there is no spontaneous breaking of $SL(2,R)$ symmetry.  In view of these, the case $\alpha= 0$ is rather special and should be discussed separately.}
It is also  suggestive for the correspondence between two pictures that the symplectic group $Sp(2)$ for the Grassmann variables is isomorphic to the $SL(2,R)$ mentioned above.  The action of the NLS model may have a wrong sign depending on the signature of the parameter $\alpha$.  This might be related to the fact that the ghost condensate does not vanish even in $g=0$ (rather diverges) for $\alpha<0$.  Note that the dimensional reduction does not imply the equivalence between two Hilbert spaces on which the respective quantum theory is constructed.
Thus the mass generation could be  related to the spontaneous breaking of $OSp(4|2)$ symmetry as claimed in \cite{KondoVI} from slightly different viewpoint.
Obviously, we need further study on the symmetry breaking.

\section{Conclusion and discussion}

We have shown that the masses of off-diagonal gluons and off-diagonal ghosts are dynamically generated in QCD by adopting the MA gauge.  This provides an evidence of the Abelian dominance which is expected to hold in low-energy region of QCD.  The MA gauge is a nonlinear gauge and hence the quartic ghost interaction term is inevitably generated by radiative corrections \cite{KondoI}.
From the viewpoint of renormalizability of the theory, therefore, we need to add the bare quartic ghost interaction to the original Lagrangian.  We have explicitly shown that the quartic ghost interaction leads to ghost--anti-ghost condensations which give the masses of the off-diagonal gluons and ghosts in QCD, although QCD doesn't have any elementary scalar field.
\par
In this Letter we determined the form of the ghost interaction from the requirement of preserving the hidden supersymmetry (the resulting gauge is called the modified MA gauge). 
 Surprisingly, the resulting Lagrangian in the modified MA gauge exactly coincides with that recently proposed by Schaden \cite{Schaden99} (at least for $SU(2)$) from quite a different point of view.
Therefore, the ghost and anti-ghost condensation can be understood as a spontaneous breaking of the global SL(2,R) symmetry recently claimed by Schaden for the SU(2) case.  We have proposed an extended BRS algebra which includes the SL(2,R) algebra.
However, it is not clear at present whether the SL(2,R) symmetry can be applied to the gauge group $SU(N)$  for $N>2$.  Finally we argued that the mass generation is also related to the spontaneous breaking of a supersymmetry hidden in the modified MA gauge for arbitrary $N$.  
\par
In this Letter, although we have pointed out the importance of the quartic interaction term from renormalizability point of view, we have not indicated that the APEGT obtained in our scenario is really renormalizable.  The totally renormalizable APEGT can be obtained improving the previous work \cite{KondoI}, see \cite{KS00}.
\par
Finally, it will be interesting to see how the dynamical mass generation just obtained affects the dual (magnetic) theory.  This issue will be discussed from APEGT in a forthcoming paper \cite{KS00}.

\section*{Acknowledgments}
After submitting this paper for publication, the authors were informed by Martin Schaden that he discussed the ghost condensation and its relation to the trace anomaly and obtained the similar results (in the SU(2) case) to ours presented in this paper.
This work is supported in part by
the Grant-in-Aid for Scientific Research from the Ministry of
Education, Science and Culture (10640249).



\baselineskip 14pt

\begin{thebibliography}{99}

\bibitem{tHooft81}
  G. 't Hooft,
  Nucl.Phys. B 190 [FS3], 455-478 (1981).

 
\bibitem{EI82}
  Z.F. Ezawa and A. Iwazaki,
  Phys. Rev. D 25, 2681-2689 (1982).

\bibitem{KLSW87}
  A. Kronfeld, M. Laursen, G. Schierholz and U.-J. Wiese,
  Phys. Lett. B 198, 516-520 (1987).  

\bibitem{SY90}
  T. Suzuki and I. Yotsuyanagi,
  Phys. Rev. D 42, 4257-4260 (1990).



\bibitem{review}
  A. DiGiacomo, 
  hep-lat/9802008;  
  hep-th/9603029.
\\
  M.I. Polikarpov,
  hep-lat/9609020.
  M.N. Chernodub and M.I. Polikarpov,
  hep-th/9710205.
  \\
  G.S. Bali,
  hep-ph/9809351.

\bibitem{KondoI}
  K.-I. Kondo,
  hep-th/9709109,
  Phys. Rev. D 57, 7467-7487 (1998).
  \\
  K.-I. Kondo, 
  hep-th/9803063,
  Prog. Theor. Phys. Supplement, No. 131, 243-255.


\bibitem{Nambu74}
  Y. Nambu,
  Phys. Rev. D 10, 4262-4268 (1974).
\\
  G. 't Hooft,
  in: High Energy Physics, edited by A. Zichichi 
(Editorice Compositori, Bologna, 1975).
\\
  S. Mandelstam,
  Phys. Report  23, 245-249 (1976).

\bibitem{KT99}
  K.-I. Kondo and Y. Taira,
  hep-th/9906129,
Mod. Phys. Lett. A, to appear; 
  hep-th/9911242.
 
  
\bibitem{KondoII}
  K.-I. Kondo, 
  hep-th/9801024,
  Phys. Rev. D 58, 105019 (1998).

\bibitem{KondoIII}
  K.-I. Kondo,
  hep-th/9803133,
  Phys. Rev. D 58, 085013 (1998).

\bibitem{KondoIV}
  K.-I. Kondo,
  hep-th/9805153,
  Phys. Rev. D 58, 105016 (1998).

\bibitem{KondoV}
  K.-I. Kondo,
  hep-th/9810167,
  Phys. Lett. B 455, 251 (1999).

\bibitem{KondoVI}
  K.-I. Kondo,
  hep-th/9904045.


\bibitem{QR98}
  M. Quandt and H. Reinhardt,
 hep-th/9707185,
 Int. J. Mod. Phys. A13, 4049-4076 (1998). 
 

  
\bibitem{KS98}
  K.-I. Kondo and T. Shinohara,
  in: Proceedings of the fall meeting of Physical Society of Japan, 3-6 October 1998, Akita Univ.

\bibitem{KS00}
  K.-I. Kondo and T. Shinohara,  Renormalizable Abelian-projected effective gauge theory derived from Quantum Chromodynamics,
hep-th/0005125, and papers in preparation.

\bibitem{antiBRST}
  G. Curci and R. Ferrari,
  Phys. Lett. B 63, 91-94 (1976).
\\
  I. Ojima,
  Prog. Theor. Phys. 64, 625-638 (1980).
  

\bibitem{MLP85}
  H. Min, T. Lee and P.Y. Pac,
  Phys. Rev. D 32, 440-449 (1985).

\bibitem{AS99}
  K. Amemiya and H. Suganuma,
  Phys. Rev. D 60, 114509 (1999).

\bibitem{Schaden99}
  M. Schaden,
  hep-th/9909011, revised version.

\bibitem{Ojima82}
  I. Ojima,
  Z. Phys. C. 13, 173-177 (1982).

\bibitem{FPS84}
  M. Flensburg, C. Peterson and L. Sk\"old,
 Z. Phys. C 22, 293 (1984).




\bibitem{Schaden98}
  M. Schaden,
hep-lat/9805020,
  Phys. Rev. D 59, 014508 (1998).
\\
  L. Baulieu and M. Schaden,
  hep-th/9601039,
  Intern. J. Mod. Phys. A 13, 985 (1998).

\bibitem{Kondo00}
  K.-I. Kondo,
  Invited talk presented at International symposium on Quantum Chromodynamics and Color Confinement -- Confinement 2000 --, Osaka, Japan, March 7-10, 2000, to appear in the proceedings (World Scientific Publishing, Singapore, 2000).

\bibitem{NO80}
  N. Nakanishi and I. Ojima,
  Z. Physik C. 6, 155-160 (1980).


\bibitem{KO79}
  T. Kugo and I. Ojima,
  Prog. Theor. Phys. Suppl. 66, 1-130 (1979).

\bibitem{PS79}  
  G. Parisi and N. Sourlas,
  Phys. Rev. Lett. 43, 744-745 (1979).

\bibitem{BT81}
  L. Bonora and M. Tonin,
  Phys. Lett. B 98, 48-50 (1981).




\end{thebibliography}
\end{document}